%% file: ms.tex
\documentclass[useAMS,usenatbib,fleqn]{mn2e}
\usepackage{amsmath}
\usepackage{array}
\usepackage{booktabs}
\usepackage{xifthen}
\usepackage{graphicx}
\usepackage{color}
\usepackage{xspace}
\usepackage{url}
\usepackage{amssymb}
\usepackage{mathtools} 
\usepackage{enumitem}
\usepackage{natbib}

\include{mnras_jdf}

\newcommand{\reb}{{\sc \tt REBOUND}\xspace}
\newcommand{\rebx}{{\sc \tt REBOUNDx}\xspace}
\newcommand{\whfast}{{\sc \tt WHFast}\xspace}
\newcommand{\ias}{{\sc \tt IAS15}\xspace}

\newcommand{\bo}[1][]{%
    \ifthenelse{\equal{#1}{}}{\mathcal{O}}{\mathcal{O}\left(#1\right)}%
}

\def\mc#1{{\mathcal{#1}}}

\title[Integrators for secularly evolving systems]{On the accuracy of symplectic integrators for secularly evolving planetary systems}
\date{Accepted 2019 October 16. Received 2019 October 16; in original form 2019 August 9.}

\author[Rein et al.]{Hanno Rein$^{1,2,3}$, 
    Garett Brown$^{1,3}$,
    Daniel Tamayo$^{4}$\thanks{NHFP Sagan Fellow}\\
$^1$ Department of Physical and Environmental Sciences, University of Toronto at Scarborough, Toronto, Ontario M1C 1A4, Canada\\
$^2$ Department of Astronomy and Astrophysics, University of Toronto, Toronto, Ontario, M5S 3H4, Canada\\
$^3$ Department of Physics, University of Toronto, Toronto, Ontario, M5S 3H4, Canada,\\
$^4$ {Department of Astrophysical Sciences, Princeton University, Princeton, New Jersey 08544, United States}\\
}

\vspace{0.5\baselineskip}

\begin{document}
\maketitle

\begin{abstract}
Symplectic integrators have made it possible to study the long-term evolution of planetary systems with direct $N$-body simulations.
In this paper we reassess the accuracy of such simulations by running a convergence test on 20~Myr integrations of the Solar System using various symplectic integrators.
We find that the specific choice of metric for determining a simulation's accuracy is important. 
Only looking at metrics related to integrals of motions such as the energy error can overestimate the accuracy of a method.
As one specific example, we show that symplectic correctors do not improve the accuracy of secular frequencies compared to the standard Wisdom-Holman method without symplectic correctors, despite the fact that the energy error is three orders of magnitudes smaller.
We present a framework to trace the origin of this apparent paradox to one term in the shadow Hamiltonian.
Specifically, we find a term that leads to negligible contributions to the energy error but introduces non-oscillatory errors that result in artificial periastron precession.
This term is the dominant error when determining secular frequencies of the system.
We show that higher order symplectic methods such as the Wisdom-Holman method with a modified kernel or the SABAC family of integrators perform significantly better in secularly evolving systems because they remove this specific term.
\end{abstract}


\section{Introduction}
\label{sec:intro}
The range of timescales present in planetary systems is truly astronomical, from a few days to billions of years.
Specialized integrators are therefore needed to accurately calculate trajectories over the entire lifetime of planetary systems.
In recent years, the development of such accurate numerical methods and the increase in computing power have made very long-term integrations possible \citep{LaskarGastineau2009}.

The mixed variable symplectic \cite{WisdomHolman1991} integrator (WH) and higher order variants thereof are one of the preferred tools for this purpose \citep{Wisdom1996,LaskarRobutel2001}.
Recently, \cite{ReinTamayoBrown2019} implemented many of these higher order symplectic integrators in the freely available \reb package. 
This present paper uses several of these methods to perform a convergence study for integrations of the Solar System.
Typically authors report the performance of a method by monitoring how well it can conserve energy, as energy should be conserved exactly in a conservative system \citep{LaskarRobutel2001,Wisdom2018,ReinTamayo2015}.
Although the energy error is a metric involving all planets, it is dominated by only a few.
We show that some of these methods, despite reducing the energy error by three orders of magnitude, do not improve the accuracy of the apsidal precession frequencies that drive the solar system's long-term evolution.
In this paper, we present a framework which lets us resolve this apparent paradox.

We introduce the notation and describe the numerical setup and algorithms in Sect.~\ref{sec:methods}.
This includes a short overview of the symplectic integrators that we use, as well as an outline of the modified Fourier transform that we use to determine the secular frequencies of the system.
The results of our convergence study of long-term integrations are presented in Sect.~\ref{sec:results}. 
We then develop a framework which lets us understand the origin of errors in symplectic methods in Sec.~\ref{sec:analysis}.
We conclude with a summary and discussion of the implications in Sec.~\ref{sec:conclusions}.

\section{Methods}
\label{sec:methods}
We use the freely available $N$-body package \reb \citep{ReinLiu2012} to perform simulations of the Solar System with all 8 planets. 
We model the planets as point masses and treat the Earth-Moon system as one particle in our simulation.
The initial conditions are taken from the NASA Horizons database, provided by the Solar System Dynamics Group of the Jet Propulsion Laboratory\footnote{\url{https://ssd.jpl.nasa.gov}}.
The effects of general relativity are approximated by including a $1/r^3$ term in the potential \citep{Nobili1986} using the implementation provided by \rebx (Tamayo et al., in prep\footnote{Code and documentation are already available at \url{https://reboundx.readthedocs.io}}).

We integrate the simulations for 20~million years.
Although the direction in which we integrate does not matter, we here integrate backwards so that we can more easily compare the secular frequencies to those obtained by \cite{La2010}.  
We use the Simulation Archive \citep{ReinTamayo2017} to record 22000 snapshots throughout each integration.
Since our goal is to compare simulations with different timesteps to very high precision, we have to ensure that we can record the snapshots at exactly the same times for all simulations without the need for any interpolation or change of the timestep during the integration.
We achieve this by choosing the timesteps such that all snapshots occur at integer multiples of the timestep \citep[the same procedure was used by][]{La2010}.
Furthermore, we count the number of timesteps taken rather than keeping track of the integration time itself.
This ensures that the round-off errors originating in adding small numbers (the timestep) to a large number (the current simulation time) do not affect the analysis. 

We use the \ias integrator, a high order Gau\ss-Radau integrator \citep{ReinSpiegel2015}, to obtain a \emph{true} solution to which we can then compare all other simulations.
It is important to point out that we only consider it to be the true solution of our model, but not the true solution of the real Solar System.
All we need for our convergence study is a well defined model that contains the most important dynamical effects determining the evolution of the Solar System.
Other simulations, in particular those by \cite{La2010}, are better representations of the real evolution of the Solar System because they use more accurate initial conditions and include important physical effects such as tidal evolution, stellar mass loss, and perturbations from minor planets.
We come back to why we have confidence that the \ias integrator indeed provides the true solution to our model later, and also comment on it in Appendix~\ref{app:ias}. 

We run simulations with the exact same initial conditions using the \ias integrator, the standard WH integrator, the WH integrator with first symplectic correctors of order 17, WHC \citep{Wisdom1996}, the WH integrator with the lazy implementer's kernel method, WHCKL \citep{Wisdom1996}, and the SABA integrator with four function evaluations and correctors using the lazy implementer's method, SABACL4 \citep{LaskarRobutel2001,ReinTamayoBrown2019}.

Next, we introduce the notation that we use for operator splitting methods and then summarize the main features of the various integrators mentioned.
The differences between the various symplectic integrators and specifically their implementation in \reb are also described in more detail in \cite{ReinTamayoBrown2019}.

\subsection{Operator Splitting Schemes}
Let us begin by defining the Poisson bracket $\{g,h\}$ of two functions $g$ and $h$  of the canonical coordinates $(q_i, p_i)$ as
\begin{eqnarray}
    \{g,h\} \equiv  \sum_i\left(  \frac{\partial g}{\partial q_i } \frac{\partial h}{\partial p_i }-  \frac{\partial g}{\partial p_i } \frac{\partial h}{\partial q_i }\right),
\end{eqnarray}
which allows us to write down Hamilton's equations as
\begin{eqnarray}
    \dot q_i = \frac{\partial H}{\partial p_i} = \{q_i,H\}\quad\quad\text{and}\quad\quad
    \dot p_i = -\frac{\partial H}{\partial q_i} = \{p_i,H\}.
\end{eqnarray}
The time derivative of any function $g$ that depends only on $p$ and $q$ can then be written succinctly as
\begin{eqnarray}
    \dot g = \{g,H\}.  \label {gH}
\end{eqnarray}
Let us further introduce $\mc L_H$, an operator which can act on any phase space function $g$.
We call this operator the Lie derivative with respect to the Hamiltonian $H$ and define it as 
\begin{eqnarray}
    \mc L_H \,g \equiv \{g,H\}. 
\end{eqnarray}
Assume that the dynamical system we are interested in has $3N$ coordinates and $3N$ momenta.
The Lie derivative allows us to write Hamilton's equation in a compact way as
\begin{eqnarray}
    \dot y =  \mc L_H \,y \label{eq:hamil}
\end{eqnarray}
where $y(t)\equiv(q_1(t),\ldots,q_{3N}(t), p_1(t),\ldots,p_{3N}(t))$ are the canonical coordinates and momenta.

Lastly, let us define the formal solution operator $\varphi^{[H]}_t(y_0)$.
This operator returns the solution to the differential equation $\dot y = \mc L_H y$ at time $t$ with initial conditions $y_0$ given at $t=0$.

Because we do in general not know how to write down this abstract solution operator, we use an operator splitting method to approximate it. 
The idea behind an operator splitting method is to rewrite the differential equation in Eq.~\ref{eq:hamil}, to which we do not know the solution, as 
\begin{eqnarray}
\dot y = \mc L_A \, y \; + \; \mc L_B \, y.
\end{eqnarray}
By virtue of the linearity of the Lie derivative, it follows that this equation is identical to  Eq.~\ref{eq:hamil} if the have $H=A+B$.
This idea of expressing $H$ as a sum of $A$ and $B$ is often referred to as \emph{splitting the Hamiltonian}. 
Let us refer to the evolution operators corresponding to the differential equations $\dot y = \mc L_A y$ and $\dot y = \mc L_B y$ as ${\varphi}^{[A]}_{t}$ and ${\varphi}^{[B]}_{t}$ respectively.
Then, we can write down a second order (leap-frog) operator splitting scheme,
\begin{eqnarray}
    \varPsi_t(y_0) = 
    \varphi^{[A]}_{t/2} \circ   
    \varphi^{[B]}_{t} \circ   
    \varphi^{[A]}_{t/2} (y_0).\label{eq:splitting}
\end{eqnarray}
As we will see below, this is a second order method which approximates the evolution operator ${\varphi}^{[H]}_{t}$.
Note that in the above notation, the $\varphi$s are operators and we apply the right most operator first.

Next, we use a formal trick to express any solution operator $\varphi^{[X]}$ in terms of its corresponding Lie derivative operator $\mc L_X$ and an exponential (defined as an infinite series in the usual way), for example,
\begin{eqnarray}
    {\varphi}^{[H]}_{t}(y_0) = \exp\left(t \mc L_{H} \right) \text{Id}(y_0)
    =
     \exp\left(t \left(\mc L_A + L_B \right)  \right) \text{Id}(y_0). \label{eq:exp1}
\end{eqnarray}
To see that this is indeed a solution to the differential equation in Eq.~\ref{eq:hamil}, note that the series expansion of the exponential provides a series of differential operators acting on $\text{Id}(y_0)$.
The series is simply the Taylor expansion of the solution ${\varphi}^{[H]}$ around $t=0$. 
We can use this formalism to also express our splitting method from Eq.~\ref{eq:splitting} in terms of exponentials of Lie derivative operators, i.e. 
\begin{eqnarray}
    \varPsi_t (y_0)= 
    \exp\left(\frac t2 \mc L_A \right)
    \exp\Big( t \mc L_B \Big)
    \exp\left(\frac t2 \mc L_A \right) \text{Id}(y_0). \label{eq:splittingmethod}
\end{eqnarray}
This allows us to use the symmetric Baker-Campbell-Hausdorff (BCH) identity to combine the three exponentials of non-commuting operators into one exponential:
\begin{eqnarray}
    \varPsi_t (y_0)= 
    \exp\Big(t \mc L_A + t \mc L_B  \nonumber
    - \frac{t^3}{24} \left[\mc L_A,\left[\mc L_A, \mc L_B \right] \right]\\
    - \frac{t^3}{12} \left[\mc L_B,\left[\mc L_A, \mc L_B \right] \right]
    + \bo(t^5) \label{eq:groebner}
    \Big) \text{Id}(y_0),
\end{eqnarray}
where the commutator $\left[\mc L_A, \mc L_B \right]$ is short hand for $\mc L_A \mc L_B - \mc L_B \mc L_A $. 

If we compare Eq.~\ref{eq:exp1} with Eq.~\ref{eq:groebner}, we can see that the differential equations which our splitting scheme is solving are not those originating from the Hamiltonian $H$.
Although the first two terms are equal, Eq.~\ref{eq:groebner} also contains an infinite series of commutators of Lie derivatives. 

Note that by virtue of the Jacobi identity, we have
\begin{eqnarray}
    \left[ \mc L_{A} , \mc L_{B} \right]g(y) = \mc L_{\{A,B\}} \, g(y).
\end{eqnarray}
This allows us to identify the commutators of Lie derivatives in Eq.~\ref{eq:groebner} as Poisson brackets of the Hamiltonians $A$ and $B$.
We can then construct a \emph{shadow Hamiltonian} $\tilde H$ which is the original Hamiltonian $H$ plus an infinite series of Poisson brackets.
For example, the two terms at order $\bo[t^3]$ in Eq.~\ref{eq:groebner} correspond to the Poisson brackets $t^2 \{A, \{A,B\}\}$ and $t^2 \{B, \{A, B\}\}$ in the shadow Hamiltonian.

Note that all of these terms contain a factor of $t^n$ with $n \ge 2$.
In the limit of $t\rightarrow 0$ all error terms are therefore converging\footnote{There is an issue of whether the series actually converges \citep[see][]{Wisdom2018}. This is not important for the discussion in this paper.} to 0.
In practice we will choose a timestep smaller than the shortest dynamical timescale in the problem.
We indicate that we take a small but finite timestep by replacing $t$ with $dt$ from now on.

\subsection{Wisdom-Holman Mapping}
The Wisdom-Holman (WH) integrator \citep{WisdomHolman1991} is using the scheme given in Eq.~\ref{eq:splitting} together with a specific splitting of the Hamiltonian $H$.
In the WH integrator, part $A$ describes the evolution of planets on Keplerian orbits around the Sun and part $B$ describes the much smaller interplanetary forces.
The exact solutions $\varphi^{[A]}$ and $\varphi^{[B]}$ of the equations of motions associated with $A$ and $B$, $\dot y = \mc L_A y$ and $\dot y = \mc L_B y$, can be calculated efficiently.
$\varphi^{[A]}$ is referred to as the Kepler step and involves solving Kepler's equation.
This can be done using a series expansion which can be truncated once machine precision is reached.
$\varphi^{[B]}$ is referred to as the kick or interaction step.
It updates the velocities of planets by evaluating the planet-planet interaction potential.

Let us estimate the size of the additional terms in the shadow Hamiltonian of the WH method.
Because the WH integrator splits the Hamiltonian in a dominant Keplerian part $A$ and a perturbation part $B$, we can introduce a dimensionless parameter to keep track of this separation of scales by formally replacing $B$ with $\epsilon B$, where $\epsilon \ll 1$.
In the Solar System, the gravitational forces from other planets are roughly a factor of $10^{-3}$ smaller than the force from the Sun, thus in this particular case $\epsilon \approx 10^{-3}$.
Using this substitution, the two terms in the shadow Hamiltonian involving $\{A, \{A,B\}\}$ and $\{B, \{A, B\}\}$ are of order $\bo[\epsilon dt^2]$ and $\bo[\epsilon^2 dt^2]$, respectively. 
Because a factor of $\epsilon$ is present in all error terms, the WH integrator performs well in weakly perturbed systems where $\epsilon$ is small.

\subsection{Wisdom-Holman Mapping with Symplectic Correctors}
The standard WH method is a second order method in~$dt$.
Symplectic methods can also be constructed to higher order in the timestep \citep{Yoshida1990}.
To improve the accuracy for a fixed~$dt$, one can thus simply choose a higher order method to remove the dominant error terms in  Eq.~\ref{eq:groebner}.
However, this requires more force evaluations per timestep and is therefore slower.
Symplectic correctors were introduced by \cite{Wisdom1996} as an alternative to improve the accuracy with effectively no overhead.
They can be understood as a clever transformation between the original problem and integrator variables which removes terms of order $\bo(\epsilon)$ from the shadow Hamiltonian. 
Because the transformations between real and mapping variables only need to be performed when outputs are needed (not every timestep), they incur negligible computational cost.
For example a third-order corrector removes the  $\{A, \{A,B\}\}$ term which is of order $\bo(\epsilon dt^2)$.
A fifth-order corrector removes all the terms up to and including those of order $\bo(\epsilon dt^4)$, and so forth. 
In this paper, we use the symplectic correctors of order 17 and refer to the method as WHC.
The now dominant term in the shadow Hamiltonian for small timesteps is $\{B, \{A, B\}\}$, which is of order $\bo[\epsilon^2 dt^2]$.

\subsection{Wisdom-Holman Mapping with a modified Kernel}
If we wish to further improve the WHC method, we have to remove the now leading order term, $\{B, \{A, B\}\}\sim\bo(\epsilon^2 dt^2)$.
It goes beyond the scope of this paper to describe in detail how \cite{Wisdom1996} achieve this using a modified kernel.
What is important for our analysis is that this new method, WHCKL, has now a leading order term of $\bo(\epsilon^2 dt^4)$ by getting rid of the  $\{B, \{A, B\}\}\}$ term.
As summarized in \cite{ReinTamayoBrown2019}, there are different ways the kernel can be implemented.
In this paper, we use the lazy implementer's method.
In comparison to the modified kick method WHCKM, it allows for the inclusion of non-Newtonian forces.
And in comparison to the composition method WHCKC, it is significantly faster \citep{ReinTamayoBrown2019}.

\subsection{SABA integrator family}
We also test the symplectic integrators of the SABA family \citep{LaskarRobutel2001}.
Specifically, we use the SABACL4 method which uses four force evaluations and one corrector per timestep \citep{ReinTamayoBrown2019}.
The correctors in SABACL4 are responsible for removing the  $\{B, \{A, B\}\}$ term in the shadow Hamiltonian.
Note that although both \cite{Wisdom1996} and \cite{LaskarRobutel2001} use the word corrector to describe parts of their integrators, the derivation, interpretation, and implementation of the correctors differ significantly. 

The SABACL4 method has a leading order error term of  $\bo(\epsilon dt^6+\epsilon^2 dt^4)$.
For small timesteps the $\bo(\epsilon^2 dt^4)$ term dominates.  
Most importantly for our later discussion, this method got rid of the $\{B, \{A, B\}\}$ error term in the shadow Hamiltonian.

\subsection{Frequency Analysis}
We follow the procedure of \cite{Laskar1988, Laskar1990, Laskar1993, Laskar2003} to determine the secular frequencies of the Solar System using a \emph{modified Fourier transform}. 
Specifically we will compute the $g$-modes of the Solar System, which are related to the periastron and eccentricity of the planets. 
A full description of the algorithm is given in the references above \citep[see also][]{Sidlichovsky1996}.
Here, we only present a short outline of the algorithm.

We start by calculating the orbital elements in Jacobi coordinates for all planets at each of our snapshots.
For simulations that use a symplectic corrector, we need to apply the correctors beforehand.
This is done by calling the \texttt{synchronize} routine in \reb. 
We then compute the complex eccentricities,
\begin{eqnarray}
    z_i(t) = e_i(t) \exp{\left(i \varpi_i(t)\right)}\quad i=1,2,\ldots,8,
\end{eqnarray}
where $e_i(t)$ and $\varpi_i(t)$ are the eccentricity and the longitude of periastron of the $i$-th planet at time $t$.

Next, we perform a linear transformation to a basis that approximately corresponds to the proper modes.
Although not absolutely necessary, this transformation makes it easier to identify the dominant secular frequencies in each planet's orbital parameters.
We use the $S$ matrix published by \cite{Laskar1990} to do the transformation to proper modes,
\begin{eqnarray}
    \mathbf{u}(t) =  S^{-1} \mathbf{z}(t). \label{eq:s}
\end{eqnarray}
Since the Solar System's secular frequencies are changing very slowly over time, this transformation should be thought of as an approximation. 
For our purpose, the precision of this approximation is not relevant as long as we use the same transformation for the entire convergence study. 

On each proper mode $u_i(t)$, we then perform a Fourier transform and find the frequency $\omega_{i, \rm fft}$ with the maximum power. 
This frequency can only be determined to within one Nyquist frequency using a standard Fourier transform.
We would need an unreasonably large number of samples ($10^{10}$) to achieve the accuracy we need for our convergence study ($10^{-10}$).
To get around this, we search in the neighbourhood of $\omega_{i, \rm fft}$ for the frequency $\omega_{i, \rm max}$ which maximizes the following scalar product
\begin{eqnarray}
    \langle e^{i \omega_{i, \rm max}\, t}, u_i(t) \rangle = \int  e^{i \omega_{i, \rm max}\, t}\; u_i(t) \, h(t) dt,
\end{eqnarray}
where $h(t)$ is the Hanning window function.
The integral is evaluated at the discrete sampling points using the trapezoidal rule.
We refer to the frequency $\omega_{i, \rm max}$ as the secular frequency $g_i$.
Whereas one can continue with an orthogonalization procedure to extract other frequencies in each proper mode, we here only determine one frequency per mode.

\section{Long term integrations of the Solar System}
\label{sec:results}
In this section we present the results of our 20~Myr integrations of the Solar System.
With these results as a motivation, we then develop a formalism to explain them in Sec.~\ref{sec:analysis}.

\subsection{Comparing our secular frequencies to previous work}
\begin{table*}
    \caption{Secular frequencies obtained by our integrations and other authors. All frequencies are given in units of arc-seconds per year. The frequencies obtained in this work have been rounded to the same number of digits as in Laskar~et~al.~(2011) for easier comparison. 
     \label{tab:freq}}
    \begin{center}
        \begin{tabular}{ lllllllllllllllll } 
            &  $g_1$ & $g_2$ & $g_3$ & $g_4$ & $g_5$ & $g_6$ & $g_7$ & $g_8$ \\\midrule
            This work & 5.59& 7.422& 17.360& 17.914& 4.257519& 28.2457& 3.088043& 0.673301\\
            \cite{La2010}, La2010a  & 5.59& 7.453& 17.368& 17.916& 4.257482& 28.2449& 3.087946& 0.673019\\
            \cite{Spalding2018}&  5.71& 7.44& 17.19& 17.76& 4.26& 28.25& 3.09 &0.67\\
        \end{tabular}
    \end{center}
\end{table*}
As a consistency check, we start by comparing the secular frequencies obtained with our \ias simulation to previous work of \cite{La2010} and \cite{Spalding2018}.
The results are listed in Table~\ref{tab:freq}.
We can reproduce the frequencies obtained by \cite{La2010} with high precision, with the largest discrepancy of about 0.5\% being in $g_2$, the frequency associated with Venus\footnote{Given the low eccentricity of Venus, it makes sense that this mode is particularly sensitive to changes in the initial conditions.}.
The frequencies of the outer planets are all reproduced to four decimal digits or better.

We do not expect to reproduce the frequencies of \cite{La2010} exactly as these authors use a different set of initial conditions, treat the Moon as a separate object, include a model of tidal dissipation as well as the effects of some minor planets. 
Whereas the treatment of these additional effects are important to accurately model the long-term evolution of the Solar System, the high level of agreement we find even without including these effects gives us confidence that our simulations are an accurate model which captures the most important dynamics in the Solar System.

A further difference between the studies is that the frequencies obtained in this work come from a 20~Myr integration; the work of \cite{La2010} use a 20~Myr integration for $g_1$ through $g_4$ and a 50~Myr integration for $g_5$ through $g_8$; and the work of \cite{Spalding2018} use a 450~Myr integration for all their frequencies.

We note that although \cite{Spalding2018} also uses the \reb integrator package and initial conditions from NASA Horizons, we were not able to reproduce their frequencies as accurately as those of \cite{La2010}.
Possible explanations for this discrepancy might be their treatment of general relativity or a difference in their data reduction.
Given that a precise integration was not necessary for the conclusions of \cite{Spalding2018} and that we agree with \cite{La2010} to such high precision, we have confidence in this being the right solution. 

\subsection{Convergence study}
\begin{figure}
    \centering
    \resizebox{0.99\columnwidth}{!}{\includegraphics[trim=0cm 0cm 0cm 0]{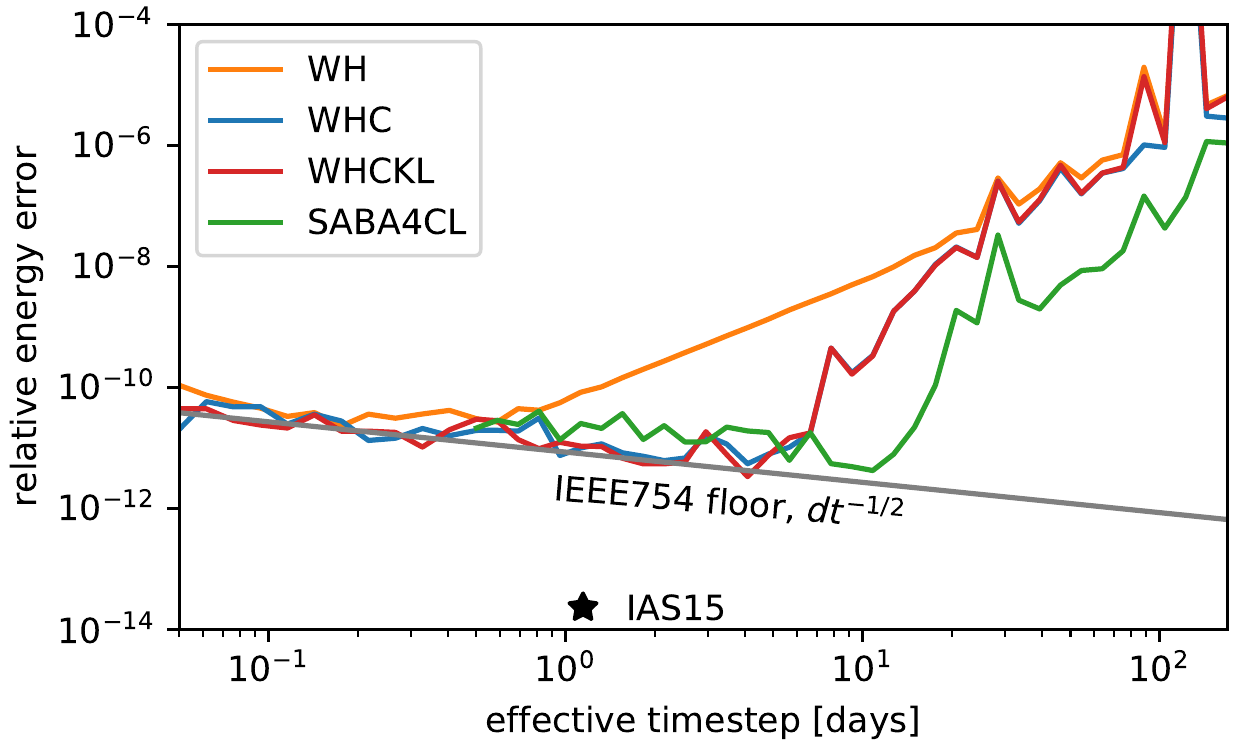}}
    \caption{Relative energy error in 20~Myr integrations of the Solar System using different integrators.
    \label{fig:longterm_energy}
    }
\end{figure}
\begin{figure*}
    \centering
    \resizebox{0.99\textwidth}{!}{\includegraphics[trim=3cm 3cm 3.4cm 2.4cm]{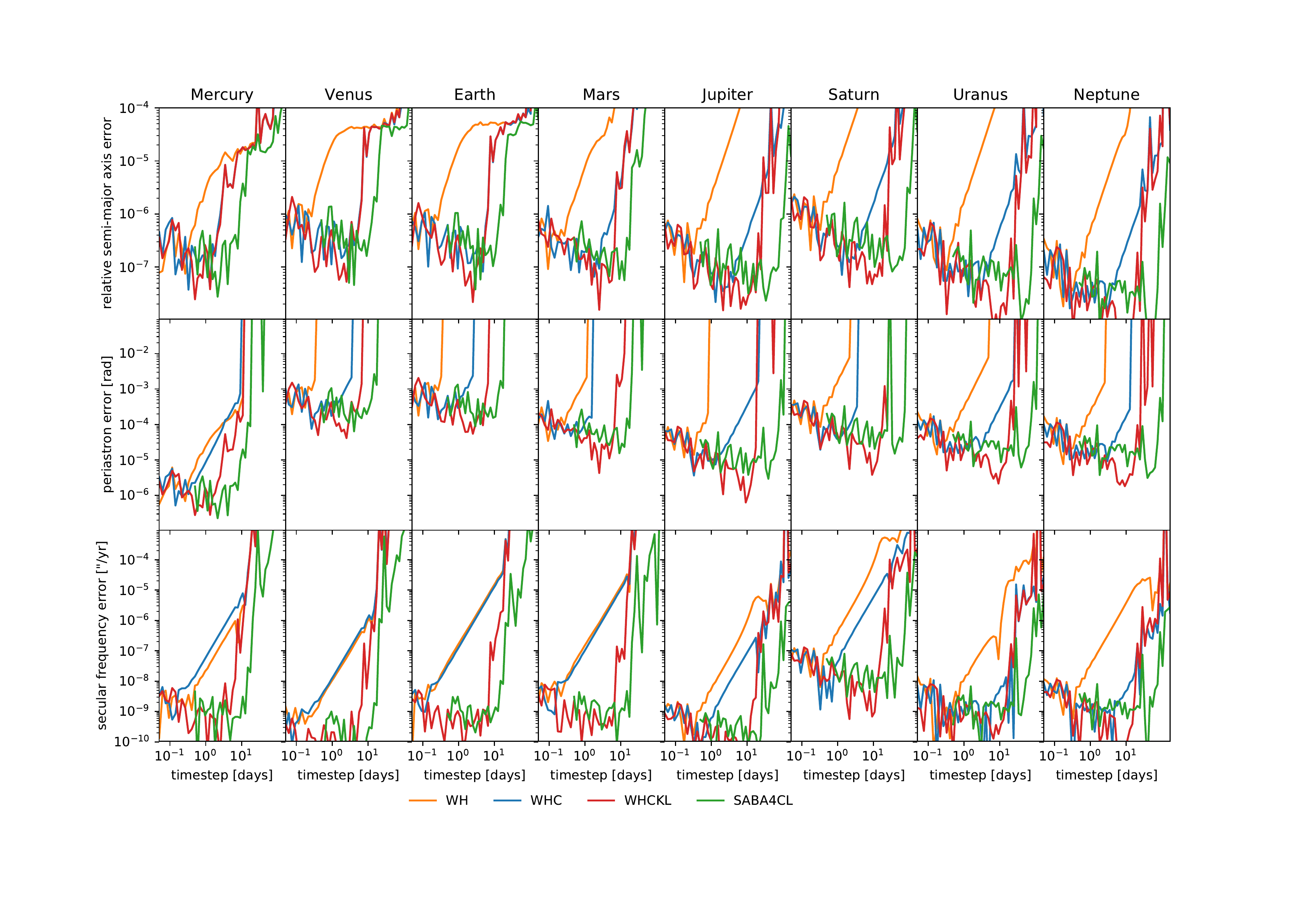}}
    \caption{Relative semi-major axis error, periastron error, and secular frequency error for all planets in 20~Myr integrations of the Solar System using different integrators.
    \label{fig:longterm_freqs}
    }
\end{figure*}

Next, we present the results of our convergence study of the WH, WHC, WHCKL, and SABACL4 integrators in our 20~Myr integrations of the Solar System.
For that, we vary the timestep from~$0.05$~days, to~$100$~days and measure various different error metrics for each simulation.
For small enough timesteps, all integrators should converge to the true solution.

Fig.~\ref{fig:longterm_energy} shows the maximum relative energy error in our simulations.
Since we are integrating a conservative system the energy should be conserved exactly by an ideal integrator and we can therefore measure the energy error directly without the need to know the exact solution.
This is why many studies only look at the energy error.
We can see that the energy error in simulations using the WH integrator converges following a $dt^2$ power-law until it reaches machine precision at $dt\approx1$~day. 
We overplot the expected  floor originating from IEEE754 double floating point round-off errors,
\begin{eqnarray}
    E_{\rm floor} \approx 2^{-53} \sqrt{N_{\rm step}}, \label{eq:brouwer}
\end{eqnarray}
where $N_{\rm step}$ is the number of timesteps required for a 20~million year integration \citep{Brouwer1937,QuinnTremaine1990}.
We cannot expect to achieve a better accuracy than that because floating point numbers on a computer are represented using a finite number of digits (about 16 in decimal, 53 in binary).
The above estimate assumes that one rounding error of $\sim 10^{-16}$ occurs after every timestep, ignoring other operations during the timestep.
Clearly, every integrator has many operations per timestep but this provides a lower limit.

The energy error in simulations using the WHC and WHCKL integrators  converge much faster and reach machine precision at $dt\approx8$~days (the two curves for WHC and WHCKL are on top of each other).
Similarly, the energy error in the SABACL4 simulations reaches machine precision at $dt\approx20$~days. 
Note that the SABACL4 integrator uses four sub-timesteps and therefore four force evaluations per timestep (plus one corrector step also requiring a force evaluation).
If we were to take this into account and plot the error as a function of the effective timestep, corresponding to one force evaluation per timestep, then the SABACL4 curve would shift to the left. 

From this plot, we might conclude that the WHC, WHCKL, and SABACL4 integrators are thus very similar in performance and several orders of magnitude better than standard WH for intermediate timesteps.

As a reference, we also show in Fig.~\ref{fig:longterm_energy} the relative energy error in the simulation using the \ias integrator. 
Since \ias is an adaptive scheme, we plot the average timestep on the horizontal axis.
Note that \ias uses compensated summation for some crucial parts of the calculation and therefore achieves an accuracy better than the IEEE754 floor.
The fact that the energy error in the \ias simulation is significantly smaller than for the other integrators gives us confidence that we can consider it the true solution of our model for all practical purposes (see also Appendix~\ref{app:ias}). 

Finally, note that the energy error is really not a particularly good metric for the convergence study in the first place.
There is only about one decade in the timestep, from $dt=10$~days to $dt=1~$day, that we can effectively use. 
For timesteps larger than 10~days, the methods are not well converged because we barely resolve the shortest timescale in the problem, Mercury's orbital period of 88~days.
For timesteps smaller than 1~day, all methods show an energy error which is dominated by round-off error, suggesting that if we further decrease the timestep, we do not increase the accuracy anymore (this conclusion is wrong as we show below). 
Part of the reason why the energy error is not a good metric for Solar System simulations is that it is a global error metric, but the Solar System has a wide range of orbital periods and a wide range of planet masses.
Jupiter's mass is more than 5000 times larger than that of Mercury, but Mercury's orbital period is 50 times shorter. 
Mercury has the fewest timesteps along its orbit, but because of its low mass, the errors in its trajectory contribute comparatively little to the total energy of the system.
Nevertheless, it is Mercury's orbit which is most likely to undergo macroscopic instabilities over billions of years \citep{LaskarGastineau2009}.

Since the energy error makes it hard to isolate potential issues, we plot different quantities in Fig.~\ref{fig:longterm_freqs}. 
Specifically, we plot the semi-major axis, periastron, and secular frequency errors associated with each planet.
Although all secular modes couple to each planet's orbital elements to some degree via Eq.~\ref{eq:s}, we associated the secular frequency $g_1$ with Mercury, $g_2$ with Venus, and so forth, following traditional conventions. 
Unlike for the energy error, we need to compare the integrations to the true solution to calculate the error in the secular frequencies.
We here consider the \ias integration as the true solution.
The semi-major axis error is scaled relative to each planet's initial semi-major axis.

Looking at the top row, we can see that the semi-major axis error for all planets follows qualitatively the results for the energy error.
The WHC, WHCKL, and SABACL4 integrators perform significantly better than the standard WH integrator.
This is expected as to leading order the energy error is simply the sum of the semi-major axes error weighted by planet mass.

However, looking at the middle and bottom row of Fig.~\ref{fig:longterm_freqs}, we can see that the periastron errors and the secular frequency errors behave very differently.
Using those metrics, the accuracy of the WHC and WH integrators are now suddenly comparable for the inner Solar System, with WHC even having a slightly larger secular frequency error in $g_1$ (bottom left panel) than WH.
The other two integrators in our sample, WHCKL and SABACL4, remain significantly better than the WH method.

This, the fact that the WHC method is performing so well when looking at the energy error, but so poorly\footnote{Relatively speaking, of course.} when looking at quantities important for the secular evolution, is the main paradox which we will explain in Sec.~\ref{sec:analysis}.

Note that the secular frequency error seems to be a particularly well suited metric for this convergence study.
We can now clearly see a second order power law for the WHC method when changing the timestep $dt$ over several orders of magnitude. 
Compare this to the energy error in Fig.~\ref{fig:longterm_energy}, where we were not able to see a power law for WHC.
Furthermore, note that the simulations which appeared to be dominated by round-off error for small timesteps in Fig.~\ref{fig:longterm_energy}, do in fact continue to improve in accuracy if the timestep is reduced further.

\subsection{High accuracy of secular frequencies}
Before we go on to derive a formalism to explain these results, let us reiterate that we use several integrators which are based on conceptually different approaches to solving the evolution of planetary systems.
\ias can be thought of as a general purpose high-order integrator, using compensated summation, variable timesteps, and not making use of the fact that we are integrating a Hamiltonian system.
Then there is the SABACL4 integrator which can also be thought of as a very general purpose scheme for nearly integrable systems, not necessarily being derived for only Hamiltonian systems \citep{LaskarRobutel2001}.
And there is the family of WH integrators whose derivation was heavily influenced by the dynamical systems approach \citep{Wisdom2018}.

Above, we have shown evidence that all these different integrators agree on the secular frequencies of the Solar System to a precision of $10^{-9}$ arc-seconds per year for small enough timesteps.
This might not be surprising. 
The integrators should converge to the same solution, after all.
But given that the integrations differ significantly and consist of up to $10^{11}$ timesteps, this is nevertheless a reassuring result.

Note that determining the secular frequencies to extremely high precision does not immediately lead to new physical insight about the system. 
In fact, the frequencies have little obvious physical meaning beyond approximately 1~part in~$10^2$ for the inner planets and 1 part in $10^4$ for the outer planets in a secular framework because they can be no longer considered constants at higher precision.
That is why they are only quoted to a finite precision in Tab.~\ref{tab:freq}.
For example, if we were to measure the frequencies in an even slightly shorter or longer integration, they would change by a factor much larger than $10^{-9}$, the precision to which we have determine them.
However, the precise physical interpretation of the secular frequencies does not matter for our convergence study. 
The way to think about the secular frequencies is not as fundamental constants of the Solar System.
Rather, they are complicated combinations of action and angle variables which are important for the long-term evolution of the system. 
And because we can measure them so precisely in a numerical simulation, we can use them to verify the behaviour of our numerical methods in an idealized but well defined model of the Solar System. 
As we have shown, the observed behaviour of some methods in numerical experiments is not always what their formal properties might suggest.

\section{Analysis}
\label{sec:analysis}
In this section we will explain why, compared to the standard WH method,  the WHC method improves the energy error but not the secular frequency error.
To do this, we need to introduce a bit of notation first.
We use an operator based approach, with a notation similar to that in \cite{ReinTamayoBrown2019} and \cite{Rein2019}.
For different perspectives, we refer the reader to books and papers by \cite{Hairer2006,LaskarRobutel2001}, and \cite{Wisdom2018} as well as references therein.

\subsection{Error operator}
We are interested in the error of our splitting methods.
Let us therefore define an error operator as
\begin{eqnarray}
    \mc E_t(y_0) \equiv {\varphi}_{t}(y_0)- \varPsi_t (y_0). 
\end{eqnarray}
At every point in phase space, this operator returns the difference between the true solution ${\varphi}$ and the approximate solution of our splitting method $\varPsi$ after some time $t$.
We can express this operator using exponentials from Eqs.~\ref{eq:exp1} and~\ref{eq:groebner}, expanding them, and then matching terms of the same power in $t$. 
This quickly becomes tedious but can be done by hand for the leading order term that we're interested in.
To leading order, we have
\begin{eqnarray}
    \mc E_t 
    &=&\nonumber 
    \Big(
        \text{Id} + t \mc L_A + t  \mc L_B 
        +\frac{t^2}{2} \left( \mc L_A +  \mc L_B \right)^2\\
    &&     +\frac{t^3}{6} \left( \mc L_A +  \mc L_B \right)^6 + \bo(t^4)\nonumber
    \Big)  \\
    &&- \Big(
        \text{Id} + t \mc L_A + t  \mc L_B \nonumber 
        +\frac{t^2}{2} \left( \mc L_A +  \mc L_B \right)^2\\
    &&     +\frac{t^3}{6} \left( \mc L_A +  \mc L_B \right)^6 \nonumber
        - \frac{t^3}{24} \left[\mc L_A,\left[\mc L_A, \mc L_B \right] \right]\\
    &&    - \frac{t^3}{12} \left[\mc L_B,\left[\mc L_A, \mc L_B \right] \right]+ \bo(t^4)
    \Big) \nonumber\\
    &=&  
         \frac{t^3}{24} \left[\mc L_A ,\left[\mc L_A, \mc L_B \right] \right]
        + \frac{t^3}{12} \left[ \mc L_B,\left[\mc L_A, \mc L_B \right] \right]
        + \bo(t^4) \label{eq:et} 
    . \label{eq:errorwh}
\end{eqnarray}
Note that all terms up to second order in $t$ cancel because we have a second order splitting $\varPsi$.
To save space  we have dropped the state on which the operator is acting on from the notation above.

\subsection{Approximating the error operator using a composition}
Next, consider the composition of operators defined by
\begin{eqnarray}
    \mc C^{AAB}_t &\equiv&  \varphi^{[A]}_{t} \circ   
     \mc C^{AB}_t\circ 
    \varphi^{[A]}_{-t} \circ   
     \mc C^{BA}_t   
\end{eqnarray}
where
\begin{eqnarray}
     \mc C^{AB}_t &\equiv& \varphi^{[A]}_{t} \circ   
    \varphi^{[B]}_{t} \circ   
    \varphi^{[A]}_{-t} \circ   
    \varphi^{[B]}_{-t}, \quad\text{and}  \nonumber\\
    \mc C^{BA}_t &\equiv& \varphi^{[B]}_{t} \circ   
    \varphi^{[A]}_{t} \circ   
    \varphi^{[B]}_{-t} \circ   
    \varphi^{[A]}_{-t}
    = \left(\mc C^{AB}_t\right)^{-1}\nonumber
\end{eqnarray}
and thus
\begin{eqnarray}
    \mc C^{AAB}_t &=& \varphi^{[A]}_{2t} \circ   
    \varphi^{[B]}_{t} \circ   
    \varphi^{[A]}_{-t} \circ   
    \varphi^{[B]}_{-t} \circ   
    \varphi^{[A]}_{-t} \nonumber\\ &&\circ  \, 
    \varphi^{[B]}_{t} \circ   
    \varphi^{[A]}_{t} \circ   
    \varphi^{[B]}_{-t} \circ   
    \varphi^{[A]}_{-t}.  \nonumber
\end{eqnarray}
Note these operators look like commutators of group elements, in contrast to the commutators of Lie derivatives (which are part of an algebra). 
After repeatedly applying the BCH formula, and ignoring terms of order $\bo(t^4)$ and higher, we can rewrite this composition as
\begin{align}
    \mc C^{AAB}_t(y_0) &=  
    \exp\Big(t^3 \left[\mc L_A,\left[\mc L_A, \mc L_B \right]\right] +\bo(t^4) \Big) \text{Id}(y_0) \nonumber \\
    &= \Big(\text{Id} + t^3 \left[\mc L_A,\left[\mc L_A, \mc L_B \right]\right] \text{Id} +\bo(t^4) \Big)(y_0). \label{eq:caab} 
\end{align}
Comparing this to Eq.~\ref{eq:et} and using the analogous operator $\mc C^{BAB}$, we can explicitly calculate the error operator to leading order in $t$ via
\begin{eqnarray}
    \mc E_t(y_0) = \Big( \mc C^{AAB}_{\alpha t}  + \mc C^{BAB}_{\beta t} - 2\cdot \text{Id}  + \bo(t^4)\Big)(y_0),
\end{eqnarray}
where $\alpha=1/\sqrt[3]{24}$ and $\beta=1/\sqrt[3]{12}$.
Note that this allows us to calculate $\mc E$ without implementing any Lie brackets.
We only need the evolution operators $\varphi^{[A]}$ and $\varphi^{[B]}$.
Furthermore, if we are only interested in one of the leading order terms in Eq.~\ref{eq:et} we can calculate each term individually. 
The calculation above can be repeated for different splitting methods.
For example, the error operator of a first order method can be expressed to leading order by the composition $\mc C^{AB}_t$.

As previously discussed in Sect.~\ref{sec:methods}, due to the Jacobi identity, we have $ \left[ \mc L_{A} , \mc L_{B} \right]g(y) = \mc L_{\{A,B\}} \, g(y)$,
which allows us to identify the commutator of Lie derivatives with a corresponding Poisson bracket of Hamiltonians $A$ and $B$.
Thus, we could also compute the operator $\mc C^{AAB}$ by first explicitly calculating the Poisson bracket $\{A, \{A, B\}\}$, then calculating, and finally solving Hamilton's equations.
In practice though, this can get tedious very quickly if calculated by hand for non-trivial Hamiltonians\footnote{An automatic differentiation algorithm could do this, but most likely not very efficiently as Lie derivatives not only contribute derivatives, but also sums over all coordinates.}.
On the other hand, the composition methods we introduced above are straightforward to implement using only two already existing operators.
Note that these compositions are similar to the compositions used by \cite{Wisdom1996} to express their symplectic correctors\footnote{Similar to how these authors construct higher order symplectic correctors, we could improve our approximation of $\mc E$ by going to higher order. 
However, since we are interested in the leading order to identify relevant terms anyway, going to higher order is not of much use for us here.}.

To summarize, the composition $\mc C^{AAB}$ that we have introduced above is a way to approximate the evolution operator $\varphi_t^{\left[\left\{A,\left\{A,B\right\}\right\}\right]}$ corresponding to evolution under the Hamiltonian $\left\{A,\left\{A,B\right\}\right\}$ to leading order in $t$ at every point in phase space $y_0$ without the need to calculate any derivatives, no matter how complicated the Hamiltonian is.
Explicitly, from Eq.~\ref{eq:caab} we have
\begin{eqnarray}
    t^3 \,\mc L_{\left\{ A,\left\{ A, B\right\}\right\}} \text{Id} (y_0)
    = \left( \mc C_t^{AAB} - \text{Id} \right) (y_0) +\bo(t^4)
\end{eqnarray}
and therefore
\begin{eqnarray}
    \varphi_t^{\left[\left\{A,\left\{A,B\right\}\right\}\right]}(y_0)
    = \frac{1}{t^2}  \mc C_t^{AAB}  (y_0) +\bo(t^2). \label{eq:phiaab}
\end{eqnarray}
Note that the factor of $t^{-2}$ acts like a denominator in a finite difference. 
If the result of this operator is dominated by round-off errors, we can easily rescale $t$ on the right hand side of Eq.~\ref{eq:phiaab} to avoid these errors.

\subsection{Using the error operator}
We can use the composition operators to approximate the evolution of any function of the phase space variables under any Poisson bracket.
Here, we want to estimate the evolution under the Poisson brackets which appear at leading order in the shadow Hamiltonian of a given splitting scheme.

For example, we can approximate the change in energy that occurs after one step with the WH method $\varPsi_t$ due to the term $[\mc L_A, [\mc L_A, \mc L_B]]$ in Eq.~\ref{eq:errorwh} as follows 
\begin{eqnarray}
    \Delta_t E(y_0) = H(y_0)- H\left(  \mc C_{\alpha t}^{AAB}  (y_0)\right) +\bo(t^2),
\end{eqnarray}
where $H$ is the Hamiltonian, now used as a function operating on the phase space variables and returning the energy as a scalar.
And $\alpha$ is a constant taking into account the constant coefficients which appear in Eq.~\ref{eq:errorwh}.
In fact we can calculate how any quantity changes due to the $[\mc L_A, [\mc L_A, \mc L_B]]$ term.
For example the error in the periastron $\varpi$ of a planet after a step of size $t$ is given by 
\begin{eqnarray}
    \Delta_t \varpi(y_0) =\varpi(y_0)-\varpi \left(\mc C_{\alpha t}^{AAB}  (y_0)\right) +\bo(t^2). \label{eq:errvarpi}
\end{eqnarray}
Compare this to simply calculating the Poisson bracket $\left\{A,\left\{A,B\right\}\right\}$ explicitly. 
This would give us the change in energy.
But to calculate the change in any other quantity, we would need the extra steps of calculating the corresponding equations of motions and then solving them.

\subsection{One planet and general relativistic precession}
\begin{figure}
    \centering
    \resizebox{0.99\columnwidth}{!}{\includegraphics[trim=0cm 0cm 0cm 0]{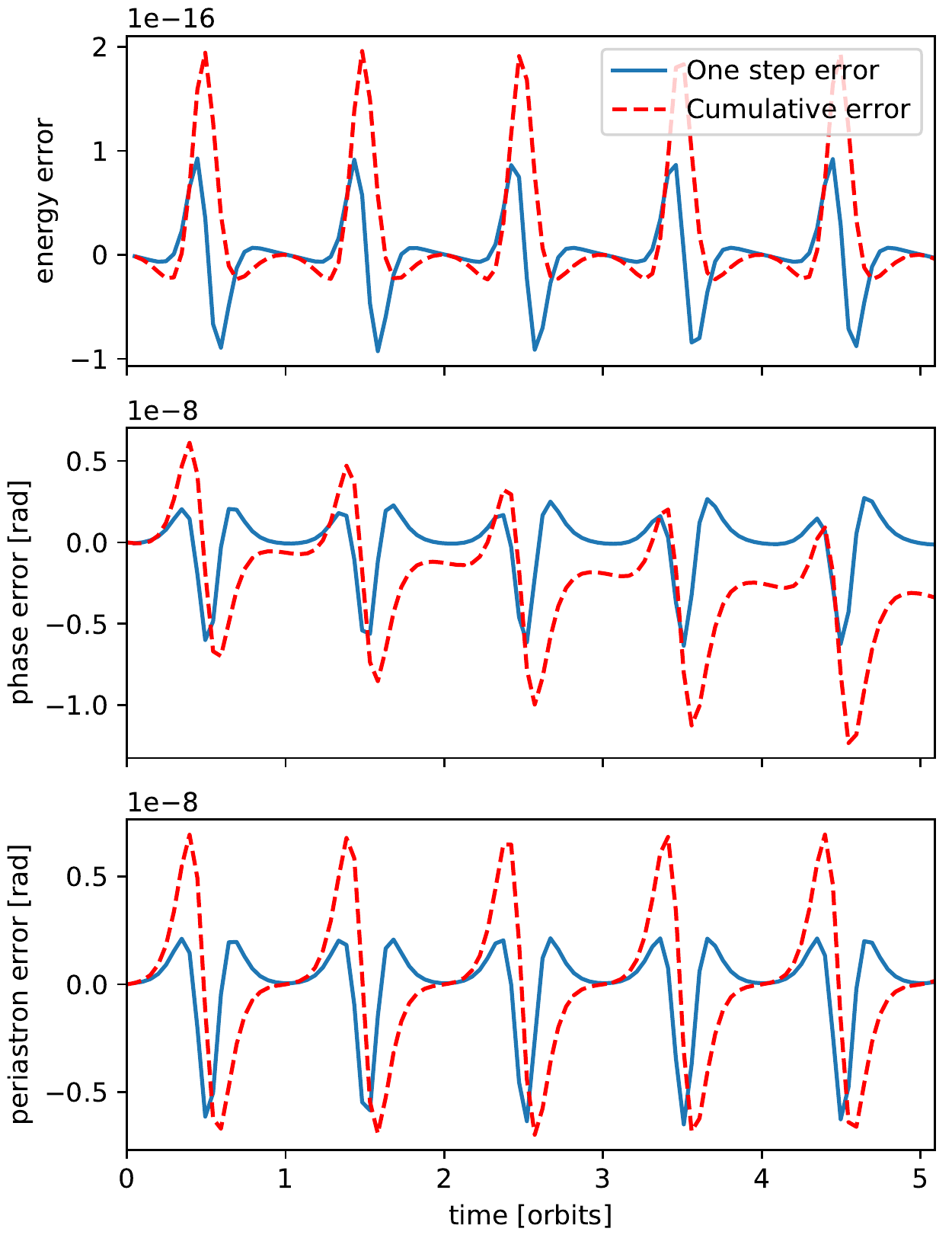}}
    \caption{Energy, phase, and periastron error in a simulation with one planet and general relativistic corrections. 
    Solid lines are one step errors, dashed lines are cumulative errors.
    \label{fig:oneplanet_error}
    }
\end{figure}
Let us start exploring how these tools might help us understand the longterm behaviour of symplectic integrators by looking at a very simple test problem consisting of just one planet orbiting a central mass  together with general relativistic corrections in the form of an additional $1/r^3$~term in the potential \citep{Nobili1986}.
We treat the general relativistic corrections as a perturbation in part $B$ of the Hamiltonian.
In this case they are the sole perturbation as there are no other planets.
The dominant Keplerian motion of the planet is described by $A$.
In the particular case we are looking at, Mercury and the Sun, the general relativistic corrections are small, leading to a precession rate of 0.5~arc-seconds per year \citep{Park2017}. 

None of the symplectic methods described above will be able to solve this problem exactly because $\varphi^{[A]}$ and $\varphi^{[B]}$ don't commute and, among others, the Poisson brackets $\{A, \{A,B\}\}$ and $\{B, \{A,B\}\}$ are non-zero. 
Let us focus on the standard WH method for now.
In Fig.~\ref{fig:oneplanet_error}, we plot the one-step as well as the cumulative errors of an integration with the WH integrator and a timestep of about 10\% of the planet's orbit. 
Note that in contrast to the one-planet Kepler problem without GR, here neither the phase nor the periastron are exact action or angle variables, but combinations thereof. 
As one can see in the first panel, both the one-step and the cumulative energy errors oscillate but show no sign of any secular increase, as one might expect for a symplectic method.
The same conclusion could be reached by looking at the periastron errors shown in the third panel, which do seem to average out, and no long-term drift is observed.
However, one can see in the middle panel, the one step phase errors do not average out, leading to the absolute value of the cumulative phase error to increase linearly with time over long timescales.
In a system such as the Solar System which evolves primarily due to secular interactions on long timescales, it is important to get the periastron resolved accurately.
The phases on the other hand matter in systems which are dominated by mean motion resonances or encounters.

Typically, one would estimate the error of the method by looking at Eq.~\ref{eq:groebner} and conclude that the leading order term is $\bo(\epsilon dt^2)$ coming from $\{A, \{A,B\}\}$.
And indeed, this does provide a good estimate of the energy error. 
However, this is not the only metric we might want to use.
For example, we might reasonably ask how the errors in the planet's other orbital elements behave.
As mentioned above, one way to estimate these errors would be to analytically calculate the evolution under $\{A, \{A,B\}\}$ and $\{B, \{A,B\}\}$.
For this simple one planet test case this is straightforward to do, at least to leading order.
But let us use the composition method introduced above which allows us to quickly estimate the errors due to the terms in  Eq.~\ref{eq:groebner} using any metric we want by using the already implemented operators~$\varphi^{[A]}$ and~$\varphi^{[B]}$.

\begin{figure}
    \centering
    \resizebox{0.99\columnwidth}{!}{\includegraphics[trim=0cm 0cm 0cm 0]{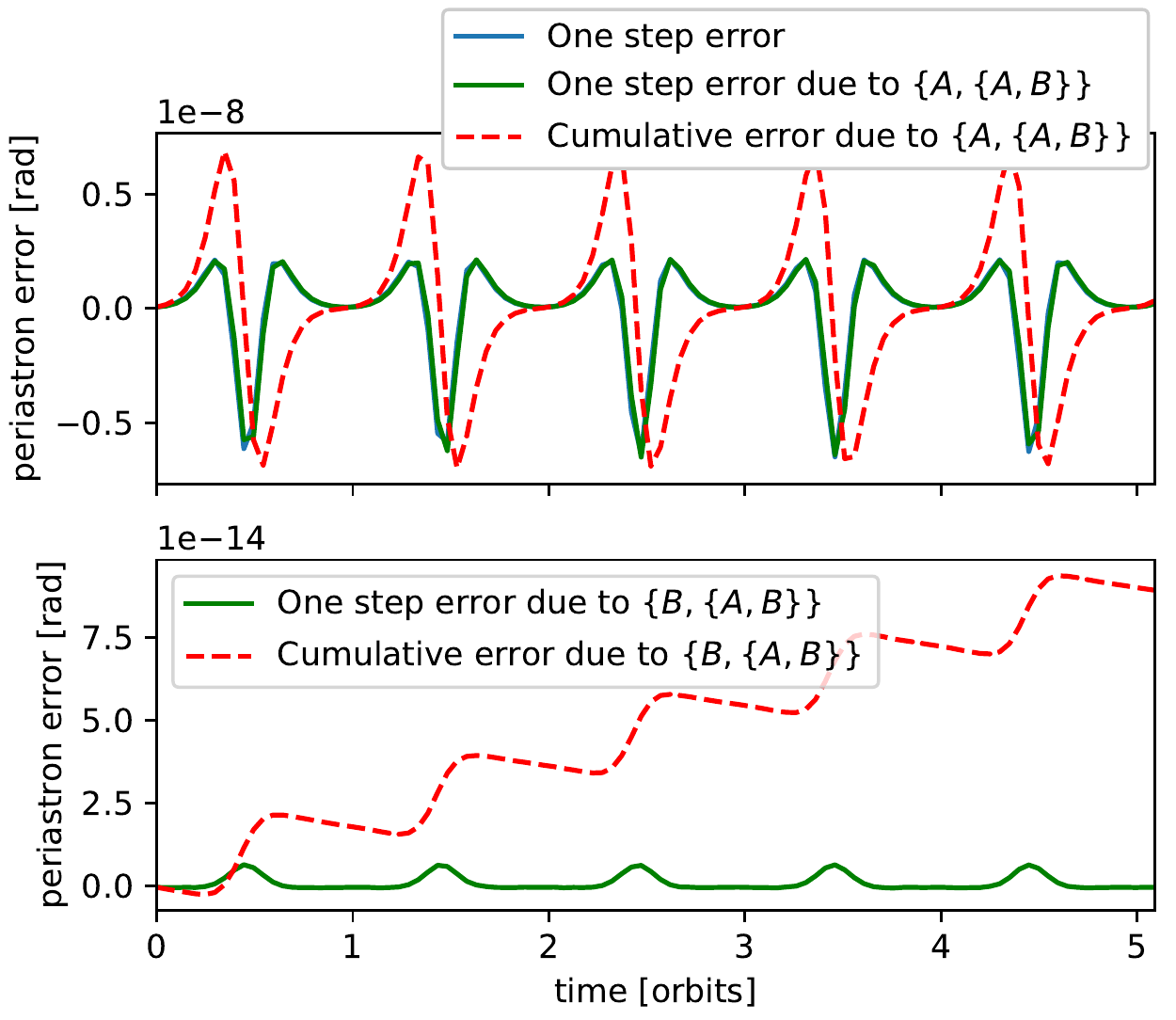}}
    \caption{Periastron error in a simulation with one planet and general relativistic corrections. 
    Solid lines are one step errors, dashed lines are cumulative errors.
    \label{fig:oneplanet_com}
    }
\end{figure}

We plot the estimate for the periastron error due to the two leading order term in Eq.~\ref{eq:groebner} in Fig.~\ref{fig:oneplanet_com}.
We calculate these estimates using the $\mc C^{AAB}$ and $\mc C^{BAB}$ operators. 
The top panel shows that the these estimates match the actual errors (Fig.~\ref{fig:oneplanet_error}) almost perfectly (the curves are hardly visible because they are on top of each other). 
The contributions from  $\mc C^{BAB}$ are significantly smaller than those  from  $\mc C^{AAB}$.
Note the different scale on the vertical axes, approximately one factor of $\epsilon$.
Nevertheless, we can observe a \emph{qualitatively} different behaviour for the errors coming from $\mc C^{BAB}$:
these errors do not cancel out over one orbital period and lead to a secular growth in the cumulative error.

We can use these results to predict the accuracy of the WH integrator over very long timescales without any further integrations. 
Not only that, we can also correctly predict the long-term behaviour of the WH integrator with symplectic correctors and the higher order methods WHCKL and SABACL4 using only the results shown in Fig.~\ref{fig:oneplanet_com}.

\begin{figure}
    \centering
    \resizebox{0.99\columnwidth}{!}{\includegraphics[trim=0cm 0cm 0cm 0]{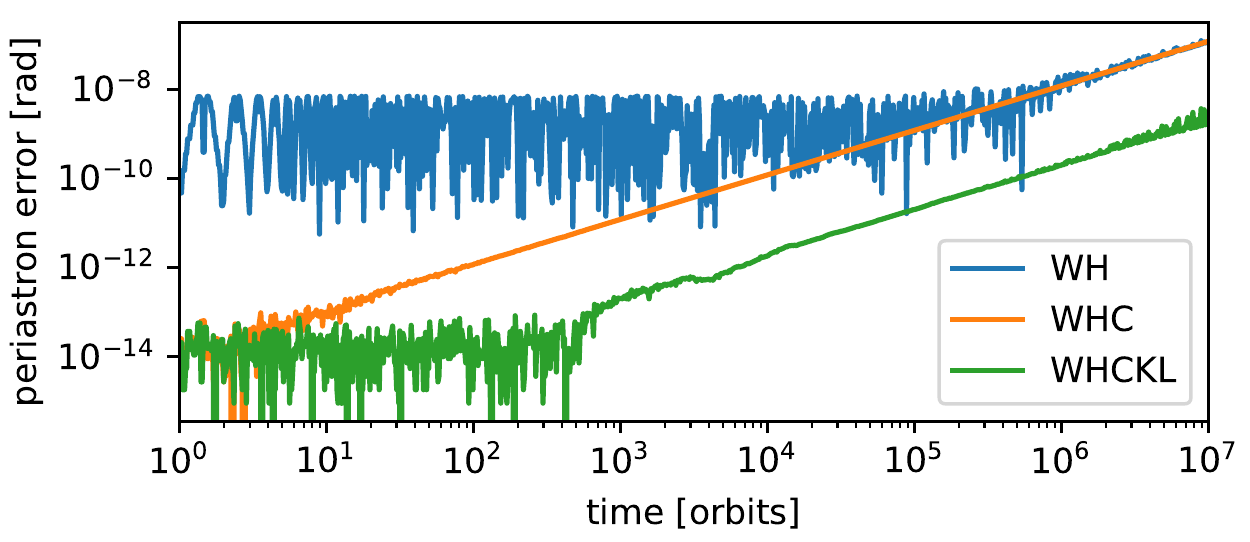}}
    \caption{Periastron error in a long-term simulation with one planet and general relativistic corrections using the WH, WHC, and WHCKL integrators.
    \label{fig:oneplanet_long}
    }
\end{figure}

Over short timescales, the secular error growth from $\{B,\{A,B\}\}$ is not important because it is so much smaller.
However, if we integrate for long enough, this will eventually become the dominant contribution to the periastron error.
Just by looking at our calculation over five orbital periods, we can estimate the growth rate and thus estimate when this happens. 
For this specific problem, the error due to $\{B,\{A,B\}\}$ will become dominant after $\approx 10^{6}$ orbital periods. 

The above calculation was done for the standard WH integrator.
However we can also use it to predict the long-term periastron error of other integrators.
If we use WHC with symplectic correctors, we remove the $\{A, \{A, B\}\}$ term.
This leads to a significant improvement of the integrations's accuracy over short timescales because the only contribution to the periastron error is now the $\{B, \{A, B\}\}$ term which is several orders of magnitude smaller (compare the top and bottom panels of Fig.~\ref{fig:oneplanet_com}).
But on long timescales, the $\{B, \{A, B\}\}$ term, which is still present in the shadow Hamiltonian even when using the WHC integrator, will grow in exactly the same way as it did for the WH integrator without correctors.
Thus, over long timescales, we do not expect any benefit from using symplectic correctors in reducing the periastron error.
We test this by running the same simulation for $10^7$~orbits and comparing the periastron error for the WH and WHC integrators to an IAS15 simulation.
The results are shown in Fig.~\ref{fig:oneplanet_long}. 
We can see that the correctors indeed improve the periastron error significantly, but only on short timescales. 
On long timescales, $\gtrapprox 10^{6}$ orbital periods, the errors are identical for the WH and WHC integrators.
Given that Mercury's orbital period is only 88 days, $10^{6}$ orbits correspond to about 250~kyrs, i.e. a short timescale compared to both the lifetime of the system and the timescale on which the dynamics of the Solar System change significantly (a few million years).

We also show the periastron error for the WHCKL integrator in Fig.~\ref{fig:oneplanet_long}.
This integrator does not have the $\{B, \{A, B\}\}$ term in the shadow Hamiltonian and we therefore expect it to perform significantly better.
A linear growth of the periastron error is still visible at late times, but this is now suppressed by two orders of magnitude.
Given that the now leading error term is two orders of magnitude smaller and the timestep is 10\% of the orbital period, we can deduce that the now leading order term must be of order $\bo[\epsilon^2 dt^4]$.

\subsection{Two planets}
\begin{figure}
    \centering
    \resizebox{0.99\columnwidth}{!}{\includegraphics[trim=2mm 0cm 1cm 0cm]{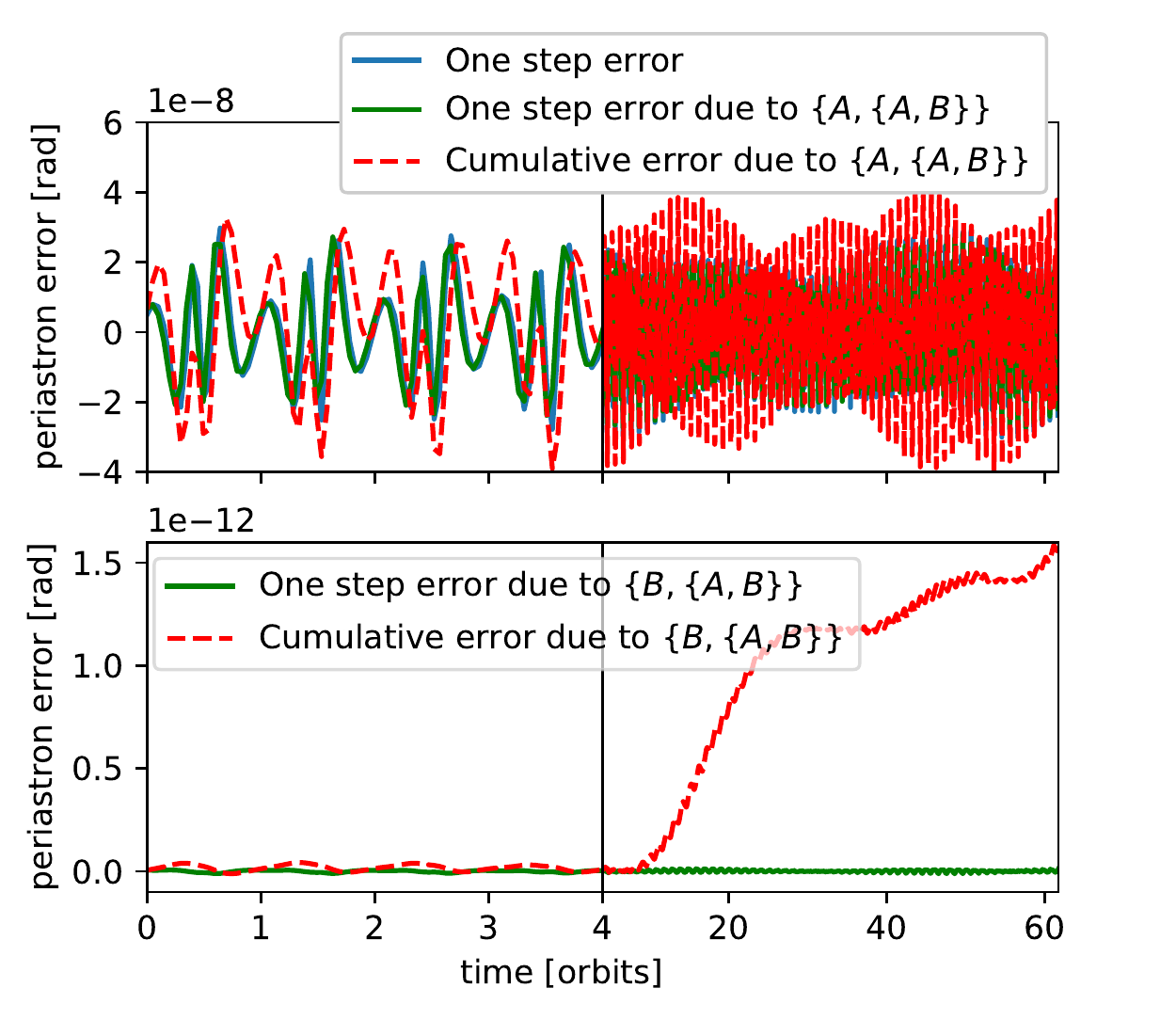}}
    \caption{Periastron error in a simulation with two planets. 
    Solid lines are one step errors, dashed lines are cumulatative errors.
    \label{fig:twoplanet_com}
    }
\end{figure}
We now repeat the line of argument from above, but this time with two planets and without general relativistic corrections.
The two planets correspond to Mercury and Saturn. 
Saturn contributes about $7.3''$~per~century to the precession rate of Mercury \citep{Park2017}.
Now part $A$ of the Hamiltonian corresponds to the Keplerian motion of the planets and part $B$ corresponds to planet-planet interactions. 

\begin{figure}
    \centering
    \resizebox{0.99\columnwidth}{!}{\includegraphics[trim=0cm 0cm 2mm 5mm]{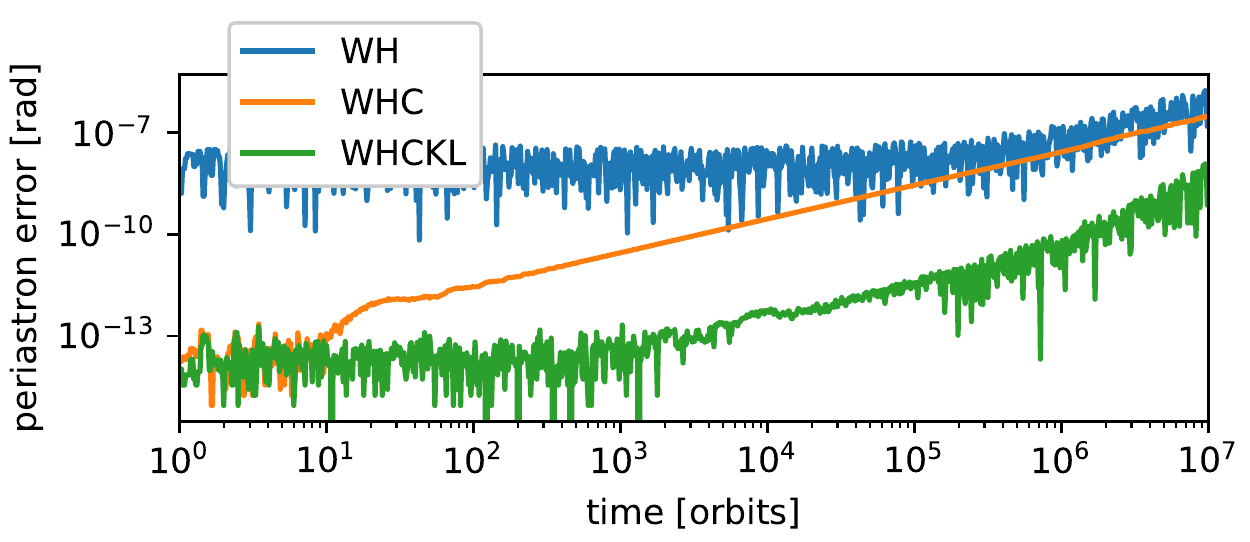}}
    \caption{Periastron error in a long-term simulation with two planets using the WH, WHC, and WHCKL integrators.
    \label{fig:twoplanet_long}
    }
\end{figure}
The same one step and cumulative errors of the periastron as before are shown in Fig.~\ref{fig:twoplanet_com}. 
And as before, we can see that the contributions to the periastron error due to $\{A,\{A,B\}\}$  average out to leading order.
The contributions due to $\{B, \{A, B\}\}$ on the other hand do not average out and we can see a secular growth, clearly visible after only 60~Mercury periods.
Thus, even though the errors due to $\{B, \{A,B\}\}$ are significantly smaller than those due to  $\{A, \{A, B\} \}$, we expect that they become dominant after $\approx 10^{5}$~Mercury periods, i.e. $25$~kyrs. 

Note that once again, we were able to predict when which error dominates without actually running a long-term integration.
To verify this, we run the two-planet case for $10^6$~Mercury period ($250$~kyr) and plot the periastron error in Fig.~\ref{fig:twoplanet_long} for simulations using the WH, WHC, and WHCKL integrators.
The results confirm our estimates. 
On short timescales, the symplectic correctors significantly improve the periastron error.
However, after about a hundred thousand years, the simulations using correctors have the same error in the periastron as those without. 
We conclude that symplectic correctors do not improve the accuracy of secularly evolving planetary systems over long timescales.

We also show the periastron error for the WHCKL integrator in Fig.~\ref{fig:twoplanet_long}.
As before, since this integrator does not have the $\{B, \{A, B\}\}$ term in the shadow Hamiltonian, it performs significantly better.
Periastron errors are still present due to higher order terms, but they are significantly suppressed.

\subsection{Implications for long-term integrations of the Solar System}
We can now come back to explaining the results of the long-term integration of the entire Solar System presented in Sect.~\ref{sec:results}. 
As compared to the simple test problems shown above, the full Solar System is a much more complex dynamical system.
Nevertheless, we can now qualitatively understand why the WHC integrator performs so poorly in reproducing the secular frequencies compared to higher order methods despite its good energy conservation properties.

This behaviour is simply a result of the $\{B, \{A, B\}\}$ term in the shadow Hamiltonian.
Although this term leads to only oscillatory behaviour in the semi-major axis error of a planet as well as the global energy error, it leads to a secular drift in the periastron error. 
If we use a timestep of $8$~days, then after only $\approx 25$~kyrs, the error in the periastron of Mercury is dominated by the cumulative contributions from the $\{B, \{A, B\}\}$ term for both the WH and WHC integrator. 
And since the periastron is used to determine the secular frequencies, we find no improvement in the secular frequencies despite the fact that other quantities such as the semi-major axes are more precise.

The integrators WHCKL and SABACL4 on the other hand get rid of the $\{B, \{A, B\}\}$ term and can therefore achieve a significantly  better accuracy of the secular frequencies. 

As one can see from Fig.\ref{fig:longterm_freqs}, the secular frequencies, and therefore the dynamics, in the outer Solar System can be improved by using symplectic correctors (bottom row).
Specifically, the secular frequencies associated with Jupiter and Saturn, $g_5$ and $g_6$, improve by about one order of magnitude.
Note that this is still significantly less than the improvement in the semi-major axes (top row) which is of order $\epsilon \sim 10^{-3}$.
The frequencies $g_7$ and $g_8$ improve even more.
We attribute this difference in the convergence of secular frequencies in the inner and outer Solar System to two effects.
First, the orbital timescales are longer in the outer Solar System. 
Thus, it will take longer for the error contributions from the $\{B, \{A, B\}\}$ term to become dominant for the outer planets.
After all, over short timescales, the correctors lead to an improvement (see Fig.~\ref{fig:twoplanet_long}) and we here only integrate for 20~Myrs. 
Second, the dynamics in the outer Solar System are also physically different from the dynamics in the inner Solar System.
For example, the semi-major axes of the outer planets change significantly ($\Delta a/a \sim \epsilon^2 \sim10^{-6}$) over the timespan of our integrations \citep{Milani1987}.
Since symplectic correctors improve the semi-major axes of all planets, we can expect them to also somewhat improve the secular frequencies in the outer Solar System if we determine them to such extreme precision as we do.
Note that this shows that secular frequencies are indeed a very useful metric to use in a convergence study given that they can capture these effects.

There is a simple physical explanation as to why removing the $\{B, \{A, B\}\}$ term is important for the accurate determination of secular frequencies, but removing the $\{A, \{A, B\}\}$ term has little effect according to our results. 
First note that the secular precession frequencies are to first order functions of the planets' semi-major axes and masses only \citep[see e.g. Exercise 7.1 in][]{solarsystemdynamics}.
However, the semi-major axes of the planets are constant to within $\bo[\epsilon]$ in a secularly evolving system as already shown by \cite{Laplace1776}.
Because the $\{A, \{A, B\}\}$ term is $\bo[\epsilon]$, its effects on the semi-major axes average out (at least to leading order in $\epsilon$).
Thus, this term does also not affect the secular frequencies over long timescales (again at least to leading order in $\epsilon$).

\section{Conclusions}
\label{sec:conclusions}
In this paper, we have shown that the energy error alone is not a sufficient metric when measuring the accuracy of long-term planetary systems which evolve due to secular interactions.
Specifically, we have presented evidence showing that the Wisdom-Holman integrator with symplectic correctors does not improve the accuracy of secular frequencies in the inner Solar System compared to the WH integrator without symplectic correctors.
This is despite the fact that the symplectic correctors improve the energy error by three orders of magnitude. 

We have presented a framework that helps us understand this behaviour.
We are now able to trace the origin of the poor performance to a specific term in the shadow Hamiltonian.
If we use a higher-order symplectic integrator such as the Kernel method of \cite{Wisdom1996} or the SABACL4 method of \cite{LaskarRobutel2001} which get rid of this term, then the integrators' performance is improved significantly. 

The use of composition operators to study a specific term in the shadow Hamiltonian has multiple advantages over an order of magnitude estimate of Poisson brackets. 
First, it lets us study terms that grow with time but are initially small and obscured by much larger terms that might not grow with time.
Second, we can study the effect of any error term on any arbitrary quantity in a simulation, i.e. not just the energy error.
This allows us to study cases where the error in the energy is oscillatory, but the error in other quantities such as the periastron error is not, leading to an increasing error with time.
The composition method is very simple to implement as it only uses already implemented functions which are part of the standard Wisdom-Holman integrator and is compatible with any position dependent potential.

We have further shown that long-term simulations of the Solar System can achieve an extremely high level of agreement in the secular frequencies across many different integrators.
This makes secular frequencies a great tool to study the accuracy of symplectic integrators.
If we use higher order methods which get rid of the dominant error terms for the periastron precession of planets, then we can achieve a remarkable level of precision. 
With a timestep of $10$~days, corresponding to only 8 timesteps per shortest dynamical timescale (Mercury's orbital period), we can calculate the secular frequency associated with Mercury to a precision of $10^{-9}$~arc-seconds per year!
However, it is important to reiterate that the purpose of this work was not an attempt to produce a list of the most accurate secular frequencies for the Solar System, but to study the performance of various symplectic integrators. 
For the former, we refer the reader to \cite{La2010} who use a much more realistic model of the Solar System than we do. 
Further, note that simply determining the secular frequencies to extremely high precision does not lead to new physical insight about the system. 
In fact, the frequencies are not well defined in secular theory beyond approximately 1~part in~$10^2$ for the inner planets and 1 part in $10^4$ for the outer planets.

Providing evidence that numerical methods converge and that physical results do not depend on nuisance parameters such as the timestep or the specific numerical integrator used are crucial for any numerical experiment. 
Most studies that check the convergence of $N$-body simulations monitor integrals of motion such as the energy.
We argue that it is wise to go beyond this and check that the quantities which are physically relevant for the evolution of the system are converged as well. 
In the case of secularly evolving systems such as the Solar System, the secular frequencies are a natural choice.
As we have shown, having an energy error which is dominated by round-off, does not necessarily imply that a simulation is converged.
Reducing the timestep even further can continue to improve the accuracy, not of the energy error of course, but of other quantities.

Finally, let us comment on how chaos fits into this picture.
It is well known that the Solar System is chaotic with the shortest Lyapunov timescale being about 5~Myr \citep{Laskar1989}.
Let us consider two of our 20~Myr simulations with two different high order integrators.
If the timestep is very small, say $dt=0.1$~days, then the simulations are converged to machine precision, $10^{-16}$.
Even if the these simulations use the exact same model and the exact same initial conditions, they will differ after only one timestep because the two integrators perform slightly different operations and this leads to a round-off error due to finite floating point precision.
Let us assume that the two simulations differ by $10^{-16}$ after one timestep.
We can then think of one of these simulations as a shadow simulation of the other, using slightly perturbed initial conditions.
If the system is chaotic, this difference will grow exponentially.
Our integration span four $e$-folding times, thus the initial difference of $10^{-16}$ will increase to about $5\cdot10^{-15}$ by the end of the simulation.
But since we introduce round-off errors at every timestep (not just in the first one) the round-off errors grow with time as well. 
By the end of the simulation the round-off error is~$\approx 10^{-10}$, see Eq.~\ref{eq:brouwer} and Fig.~\ref{fig:longterm_energy}.
We obtain the secular frequencies to a precision of $10^{-9}$.
Thus, the frequencies are clearly not affected by chaos. 
The precision is either limited by round-off errors for small timesteps, or discretization errors coming from the integrator for large timesteps as shown in Fig.~\ref{fig:longterm_freqs}.
But this is only true for this specific simulation. 
If we integrate for much longer, then the precision to which we can obtain the secular frequencies will eventually be limited by chaos.
To summarize, there are four different contributions to the divergence of trajectories in our simulations with symplectic integrators:
\begin{enumerate}[leftmargin=*]
    \item Discretization errors in the form of oscillating error terms which average out over time and lead to a constant error, $\bo[1]$.  
    \item  Discretization errors where the errors do not average out, growing linearly with time, $\bo[t]$. 
    \item Floating point round-off errors, growing like a random walk in predominantly action-like variables in an unbiased implementation, $\bo[\sqrt t]$, and growing linearly in a biased implementation, $\bo[t]$.
    For predominantly angle-like variables this can also lead to errors growing as $\bo[t^{1.5}]$ or $\bo[t^2]$, respectively.
    \item Chaos, with the seed coming from round-off errors, leading to an exponential divergence, $\bo[e^t]$. 
\end{enumerate}
Which contribution dominates depends on the system at hand, the integrator and timestep used, and last but not least the length of the integration because the various contributions depend very differently on time, i.e. $\bo[1]$, $\bo[\sqrt t]$, $\bo[t]$, $\bo[t^{1.5}]$, $\bo(t^2)$, $\bo[e^t]$.
The tools described in this paper help to determine which of these contributions is limiting the accuracy of trajectories in a specific $N$-body simulation, and in some cases, how to improve the accuracy.

\section*{Acknowledgments}
We thank an anonymous referee for helpful comments. 
This research has been supported by the NSERC Discovery Grant RGPIN-2014-04553 and the Centre for Planetary Sciences at the University of Toronto Scarborough.
Support for this work was provided by NASA through the NASA Hubble Fellowship grant HST-HF2-51423.001-A awarded  by  the  Space  Telescope  Science  Institute,  which  is  operated  by  the  Association  of  Universities  for  Research  in  Astronomy,  Inc.,  for  NASA,  under  contract  NAS5-26555.
This research was made possible by the open-source projects 
\texttt{Jupyter} \citep{jupyter}, \texttt{iPython} \citep{ipython}, 
and \texttt{matplotlib} \citep{matplotlib, matplotlib2}.

\appendix
\section{Taking IAS15 as the true solution}
\label{app:ias}
Let us comment on why Fig.~\ref{fig:longterm_freqs} provides further evidence that the \ias simulation is indeed much more accurate than any of the other simulation in our sample.
This is not to say that the \ias algorithm is superior. 
The difference is partly due to the fact that \ias uses compensated summation to reduce round-off errors whereas the other integrators do not have this capability in our implementation (see also \citealt{LaskarRobutel2001} and \citealt{Wisdom2018} who use these algorithms with some form of extended precision).

Note that all error metrics for all integrators in Fig.~\ref{fig:longterm_freqs} are eventually dominated by round-off errors at small timesteps.
Specifically, note that the curves are noisy as expected from a random walk.
Because we are comparing all simulations shown in the plot to the same \ias simulation, this noisy pattern would not show up if the \ias simulation were less accurate than the \whfast simulations, or if the round-off error came from the \ias simulation. 
If the \ias simulation had a larger error than the other simulations, then all the curves for the different integrators would be on top of each other on the left side of Fig.~\ref{fig:longterm_energy} and~\ref{fig:longterm_freqs}. 

This gives us further confidence that our assumption of using the \ias simulation as the true solution is appropriate.
Alternatively, we could have used one of the higher order symplectic integrations with the smallest timestep as the reference simulation. 
This would still allow us to determine when the error is dominated by round-off errors.
But we would not be able to determine the power law in the round-off error dominated regime.

\bibliography{full}

\end{document}

%% file: mnras_jdf.tex
\def \aap{A\&A}

\def \apjl{ApJ}
\def \apjs{ApJS}

%% file: ms.bbl
\begin{thebibliography}{32}
\expandafter\ifx\csname natexlab\endcsname\relax\def\natexlab#1{#1}\fi

\bibitem[{{Brouwer}(1937)}]{Brouwer1937}
{Brouwer}, D. 1937, \aj, 46, 149

\bibitem[{Droettboom {et~al.}(2016)Droettboom, Hunter, Caswell, Firing,
  Nielsen, Elson, Root, Dale, Lee, Seppänen, McDougall, Straw, May, Varoquaux,
  Yu, Ma, Moad, Silvester, Gohlke, Würtz, Hisch, Ariza, Cimarron, Thomas,
  Evans, Ivanov, Whitaker, Hobson, mdehoon, \& Giuca}]{matplotlib2}
Droettboom, M., Hunter, J., Caswell, T.~A., Firing, E., Nielsen, J.~H., Elson,
  P., Root, B., Dale, D., Lee, J.-J., Seppänen, J.~K., McDougall, D., Straw,
  A., May, R., Varoquaux, N., Yu, T.~S., Ma, E., Moad, C., Silvester, S.,
  Gohlke, C., Würtz, P., Hisch, T., Ariza, F., Cimarron, Thomas, I., Evans,
  J., Ivanov, P., Whitaker, J., Hobson, P., mdehoon, \& Giuca, M. 2016,
  matplotlib: matplotlib v1.5.1

\bibitem[{Hairer {et~al.}(2006)Hairer, Lubich, \& Wanner}]{Hairer2006}
Hairer, E., Lubich, C., \& Wanner, G. 2006, Geometric numerical integration:
  structure-preserving algorithms for ordinary differential equations, Vol.~31
  (Springer Science \& Business Media)

\bibitem[{Hunter(2007)}]{matplotlib}
Hunter, J.~D. 2007, Computing In Science \& Engineering, 9, 90

\bibitem[{Kluyver {et~al.}(2016)Kluyver, Ragan-Kelley, P{\'e}rez, Granger,
  Bussonnier, Frederic, Kelley, Hamrick, Grout, Corlay, {et~al.}}]{jupyter}
Kluyver, T., Ragan-Kelley, B., P{\'e}rez, F., Granger, B., Bussonnier, M.,
  Frederic, J., Kelley, K., Hamrick, J., Grout, J., Corlay, S., {et~al.} 2016,
  Positioning and Power in Academic Publishing: Players, Agents and Agendas, 87

\bibitem[{Laplace(1776)}]{Laplace1776}
Laplace, P.-S. 1776, M{\'e}moires de l’Acad{\'e}mie Royale des Sciences de
  Paris, 7, 69

\bibitem[{Laskar(1988)}]{Laskar1988}
Laskar, J. 1988, Astronomy and Astrophysics, 198, 341

\bibitem[{{Laskar}(1989)}]{Laskar1989}
{Laskar}, J. 1989, \nat, 338, 237

\bibitem[{Laskar(1990)}]{Laskar1990}
Laskar, J. 1990, Icarus, 88, 266

\bibitem[{Laskar(1993)}]{Laskar1993}
---. 1993, Physica D: Nonlinear Phenomena, 67, 257

\bibitem[{Laskar(2003)}]{Laskar2003}
---. 2003, arXiv:math/0305364

\bibitem[{{Laskar} {et~al.}(2011){Laskar}, {Fienga}, {Gastineau}, \&
  {Manche}}]{La2010}
{Laskar}, J., {Fienga}, A., {Gastineau}, M., \& {Manche}, H. 2011, \aap, 532,
  A89

\bibitem[{{Laskar} \& {Gastineau}(2009)}]{LaskarGastineau2009}
{Laskar}, J. \& {Gastineau}, M. 2009, \nat, 459, 817

\bibitem[{Laskar \& Robutel(2001)}]{LaskarRobutel2001}
Laskar, J. \& Robutel, P. 2001, Celestial Mechanics and Dynamical Astronomy,
  80, 39

\bibitem[{{Milani} {et~al.}(1987){Milani}, {Nobili}, \& {Carpino}}]{Milani1987}
{Milani}, A., {Nobili}, A.~M., \& {Carpino}, M. 1987, \aap, 172, 265

\bibitem[{{Murray} \& {Dermott}(2000)}]{solarsystemdynamics}
{Murray}, C.~D. \& {Dermott}, S.~F. 2000, {Solar System Dynamics} (Cambridge
  University Press)

\bibitem[{{Nobili} \& {Roxburgh}(1986)}]{Nobili1986}
{Nobili}, A. \& {Roxburgh}, I.~W. 1986, in IAU Symposium, Vol. 114, Relativity
  in Celestial Mechanics and Astrometry. High Precision Dynamical Theories and
  Observational Verifications, ed. J.~{Kovalevsky} \& V.~A. {Brumberg},
  105--110

\bibitem[{Park {et~al.}(2017)Park, Folkner, Konopliv, Williams, Smith, \&
  Zuber}]{Park2017}
Park, R.~S., Folkner, W.~M., Konopliv, A.~S., Williams, J.~G., Smith, D.~E., \&
  Zuber, M.~T. 2017, The Astronomical Journal, 153, 121

\bibitem[{P\'erez \& Granger(2007)}]{ipython}
P\'erez, F. \& Granger, B.~E. 2007, Computing in Science and Engineering, 9, 21

\bibitem[{{Quinn} \& {Tremaine}(1990)}]{QuinnTremaine1990}
{Quinn}, T. \& {Tremaine}, S. 1990, \aj, 99, 1016

\bibitem[{{Rein} {et~al.}(2019{\natexlab{a}}){Rein}, {Hernandez}, {Tamayo},
  {Brown}, {Eckels}, {Holmes}, {Lau}, {Leblanc}, \& {Silburt}}]{Rein2019}
{Rein}, H., {Hernandez}, D.~M., {Tamayo}, D., {Brown}, G., {Eckels}, E.,
  {Holmes}, E., {Lau}, M., {Leblanc}, R., \& {Silburt}, A. 2019{\natexlab{a}},
  \mnras, 485, 5490

\bibitem[{{Rein} \& {Liu}(2012)}]{ReinLiu2012}
{Rein}, H. \& {Liu}, S.-F. 2012, \aap, 537, A128

\bibitem[{{Rein} \& {Spiegel}(2015)}]{ReinSpiegel2015}
{Rein}, H. \& {Spiegel}, D.~S. 2015, \mnras, 446, 1424

\bibitem[{{Rein} \& {Tamayo}(2015)}]{ReinTamayo2015}
{Rein}, H. \& {Tamayo}, D. 2015, MNRAS, 452, 376

\bibitem[{{Rein} \& {Tamayo}(2017)}]{ReinTamayo2017}
---. 2017, MNRAS, 467, 2377

\bibitem[{{Rein} {et~al.}(2019{\natexlab{b}}){Rein}, {Tamayo}, \&
  {Brown}}]{ReinTamayoBrown2019}
{Rein}, H., {Tamayo}, D., \& {Brown}, G. 2019{\natexlab{b}}, arXiv e-prints,
  arXiv:1907.11335

\bibitem[{{\v{S}}idlichovsk{\'y} \& Nesvorn{\'y}(1996)}]{Sidlichovsky1996}
{\v{S}}idlichovsk{\'y}, M. \& Nesvorn{\'y}, D. 1996, Celestial Mechanics and
  Dynamical Astronomy, 65, 137

\bibitem[{{Spalding} {et~al.}(2018){Spalding}, {Fischer}, \&
  {Laughlin}}]{Spalding2018}
{Spalding}, C., {Fischer}, W.~W., \& {Laughlin}, G. 2018, \apj, 869, L19

\bibitem[{{Wisdom}(2018)}]{Wisdom2018}
{Wisdom}, J. 2018, \mnras, 474, 3273

\bibitem[{{Wisdom} \& {Holman}(1991)}]{WisdomHolman1991}
{Wisdom}, J. \& {Holman}, M. 1991, \aj, 102, 1528

\bibitem[{{Wisdom} {et~al.}(1996){Wisdom}, {Holman}, \& {Touma}}]{Wisdom1996}
{Wisdom}, J., {Holman}, M., \& {Touma}, J. 1996, Fields Institute
  Communications, Vol.~10, p.~217, 10, 217

\bibitem[{{Yoshida}(1990)}]{Yoshida1990}
{Yoshida}, H. 1990, Physics Letters A, 150, 262

\end{thebibliography}
